\documentclass[aps,prd,twocolumn,amsmath,amsfonts,amssymb,groupedaddress,nofootinbib,nobalancelastpage,floatfix,secnumarabic,superscriptaddress,secnumarabic,preprintnumbers]{revtex4-2}
\pdfoutput=1
\usepackage{epsfig, graphicx, hyperref, slashed,xspace}
\usepackage[dvipsnames]{xcolor}
\usepackage[utf8]{inputenc}
\usepackage[T1]{fontenc}
\usepackage{cleveref}
\usepackage{soul}
\usepackage{diagbox}
\usepackage{booktabs}
\usepackage{multirow}
\usepackage{setspace}
\usepackage{float}
\usepackage{subcaption}

\renewcommand{\arraystretch}{1.25}   

\hypersetup{colorlinks=true,linkcolor=Maroon,citecolor=ForestGreen,filecolor=ForestGreen,urlcolor=ForestGreen}

\newcommand{\pythiaN}{{\sc Pythia8}\xspace}
\newcommand{\pythia}{{\sc Pythia6}\xspace}
\newcommand{\whizard}{{\sc Whizard3}\xspace}

\newcommand{\ab}{{$\rm{ab}^{-1}$}\xspace}

\makeatletter\def\l@subsubsection#1#2{}\makeatother
\usepackage{orcidlink}

\begin{document}


\title{Prospects for Measuring $H\to \rm{invisble}$ at the FCCee}

\author{Aman Desai \orcidlink{0000-0003-2631-9696}}\email{aman.desai@adelaide.edu.au}
\affiliation{Department of Physics, Adelaide University, North Terrace, Adelaide, SA 5005, Australia}

\author{Paul Jackson \orcidlink{0000-0002-0847-402X}}\email{p.jackson@adelaide.edu.au}
\affiliation{Department of Physics, Adelaide University, North Terrace, Adelaide, SA 5005, Australia}

\date{\today}

\begin{abstract}

We present the prospects for measuring $H\to \rm{invisble}$ decays at the Future Circular Collider electron-positron at $\sqrt{s} = 240 \text{ GeV}$ with an integrated luminosity of 10.8 ab$^{-1}$. In this study, we consider the $ZH$ production mode with three decay modes of the $Z$ boson: $Z\to e^+e^-$, $Z\to \mu^+\mu^-$ and $Z\to jj$ ($b\bar{b}, c\bar{c}, s\bar{s}, q\bar{q}$). We find that at 95\% confidence limit, the combined upper limit on the $\mathcal{B}(H\to invisible)$ could reach 0.15\%. 
 
\end{abstract}

\maketitle

\section{Introduction}\label{sec:intro}

In the Standard Model of particle physics (SM) the Higgs boson can decay invisibly via $H\to ZZ\to \nu\bar{\nu}\nu\bar{\nu}$ with a branching fraction of 0.106\%~\cite{LHCHiggsCrossSectionWorkingGroup:2016ypw}. The ATLAS and CMS collaborations have investigated this decay channel and have reported upper limits on the branching fraction~\cite{ATLAS:2023tkt, CMS:2023sdw}.

The Future Circular Collider-ee (FCC-ee) is a proposed electron-positron collider to be built at CERN~\cite{FCC:2025lpp}. This collider facility aims to explore interactions across a centre-of-mass energy range extending from the $Z$ mass (91 GeV)  and reaching up to the top quark pair production threshold (365 GeV). Among the centre-of-mass energies, $\sqrt{s} = 240 \text{ GeV}$ is crucial for investigating the Higgs boson owing to the high cross-section for the primary production process $ZH$ at this energy~\cite{Selvaggi:2025kmd}. At these energies and with an expected integrated luminosity of  10.8 \ab, it is expected that the FCC-ee will produce $\sim 2.2 \times 10^6$ Higgs boson. This would allow precision measurements of several Higgs decay modes along with the Higgs boson's total decay width.  At the FCC-ee, at $\sqrt{s}=240$ GeV, we expect to see $\sim 10^3$ invisibly decaying Higgs boson. 

Previous studies investigating the $\mathcal{B}(H\to invisible)$ channel have been carried out in the context of the proposed lepton colliders such as the CEPC~\cite{Tan:2020ufz}, CLIC~\cite{Mekala:2020zys}, ILC~\cite{Ishikawa:2019uda} as well as the FCC-ee~\cite{mehta_2025}, among others.

We analyze the $e^+ e^- \to ZH$ process considering the Z boson decays: $Z\to qq, ee, \mu\mu$  and the Higgs boson decaying invisibly. We use multivariate analysis methods to discriminate the signal from background in addition to using a set of selection criteria. 

This paper is organized as follows: We discuss the Monte Carlo simulation in \autoref{sec:eventsim}; in \autoref{sec:analysis} we present the analysis procedure and the statistical inference is presented in \autoref{sec:stat}.

\section{Event Simulation}\label{sec:eventsim}

In this analysis, we use Monte Carlo samples prepared centrally by the FCC-ee as part of the \textsc{Winter2023} Campaign~\cite{FCC_EE_IDEA_Winter2023}. The signal process in this analysis is the production mode: $e^+ e^- \to ZH$ with $Z$ boson decaying either to a pair of electrons/muons or quarks and the Higgs boson decaying invisibly. The invisible decays of the Higgs boson are simulated via the process $H\to ZZ \to \nu\bar{\nu}\nu\bar{\nu}$.  Therefore, there are three different signals which are analyzed: $Z(ee) H(inv)$, $Z(\mu\mu) H(inv)$ and $Z( qq) H(inv)$. The list of signal processes used in this analysis and their cross-sections are given in \autoref{tab:signal_xsec_nevents}.

Physics processes that mimic the signal or have the potential to do so are considered as backgrounds. In particular, the backgrounds differ in each of the $Z$ decay channel. The set of backgrounds for three analysis is given in \autoref{tab:background_xsec_nevents} along with the corresponding cross-section and number of events simulated for analysis.

\begin{table*}[tbp]
\centering
\caption{Signal processes considered at $\sqrt{s}=240~\mathrm{GeV}$ along with the number of generated events and cross-sections~\cite{FCC_EE_IDEA_Winter2023}.}
\label{tab:signal_xsec_nevents}
\renewcommand{\arraystretch}{1.3}
\begin{tabular}{l c c c}
\toprule
\textbf{Process} & $N_{\text{events}}$ & \textbf{Cross-section [fb]} & Channel \\
\midrule
$e^+e^-  \to e^+e^- H(\to ZZ) ~\text{Inv.}$ & $1.20 \times 10^6$ & $7.52 \times 10^{-3}$ & $ee$ \\
$e^+e^-  \to \mu^+\mu^- H(\to ZZ) ~\text{Inv.}$ & $1.20 \times 10^6$ & $7.10 \times 10^{-3}$ & $\mu\mu$ \\
$e^+e^-  \to qq H(\to ZZ) ~\text{Inv.}$ & $1.20 \times 10^6$ & $5.60 \times 10^{-2}$ & $jj$ \\
$e^+e^-  \to bb H(\to ZZ) ~\text{Inv.}$ & $1.172 \times 10^6$ & $3.15 \times 10^{-2}$ & $jj$ \\
$e^+e^-  \to cc H(\to ZZ) ~\text{Inv.}$ & $1.20 \times 10^6$ & $2.45 \times 10^{-2}$ & $jj$ \\
$e^+e^-  \to ss H(\to ZZ) ~\text{Inv.}$ & $1.20 \times 10^6$ & $3.15 \times 10^{-2}$ & $jj$ \\
\bottomrule
\end{tabular}
\end{table*}

\begin{table*}[tbp]
\centering
\caption{Background processes considered at $\sqrt{s}=240~\mathrm{GeV}$ along with the number of generated events and cross-sections~\cite{FCC_EE_IDEA_Winter2023}.}
\label{tab:background_xsec_nevents}
\renewcommand{\arraystretch}{1.3}
\begin{tabular}{l c c c}
\toprule
\textbf{Process} & $N_{\text{events}}$ & \textbf{Cross-section [fb]} & Channel \\
\midrule
$e^+e^- \to ZZ$ & $5.62 \times 10^{7}$ & $1.36 \times 10^{3}$ & $ee, \mu\mu, jj$ \\
$e^+e^- \to W^+W^-$ & $3.73 \times 10^{8}$ & $1.64 \times 10^{4}$ & $ee, \mu\mu, jj$ \\
$e^+e^- \to \nu\bar{\nu} H(\to ZZ)$ No Inv. & $1.20 \times 10^{6}$ & $1.17$ & $ee, \mu\mu, jj$ \\
$e^+e^- \to \nu\bar{\nu} H(\to WW)$ & $1.20 \times 10^{6}$ & $9.94$ & $ee, \mu\mu, jj$ \\
$e^+e^- \to \nu\bar{\nu} Z$ & $2.00 \times 10^{6}$ & $33.2$ & $ee, \mu\mu, jj$ \\
$e^+e^- \to Z/\gamma^*  \to e^+e^-$ & $5.34 \times 10^{7}$ & $5.28 \times 10^{3}$ & $ee$ \\
$e^+e^- \to e^+e^- H(\to ZZ)$ No Inv. & $1.20 \times 10^{6}$ & $0.18$ & $ee$ \\
$e^+e^- \to Z/\gamma^* \to \mu^+\mu^-$ & $8.54 \times 10^{7}$ & $8.35 \times 10^{3}$ & $\mu\mu$ \\
$e^+e^- \to \mu^+\mu^- H(\to ZZ)$ No Inv. & $1.20 \times 10^{6}$ & $0.17$ & $\mu\mu$ \\
$e^+e^- \to \nu\bar{\nu} H(\to \mu\mu)$ & $4.0 \times 10^{5}$ & $0.01$ & $\mu\mu$ \\
$e^+e^- \to Z/\gamma^* \to q\bar{q}$ & $1.01 \times 10^{8}$ & $5.26 \times 10^{4}$ & $jj$ \\
$e^+e^- \to qq H(\to ZZ)$ No Inv. & $1.20 \times 10^{6}$ & $1.35$ & $jj$ \\
$e^+e^- \to bb H(\to ZZ)$ No Inv. & $1.20 \times 10^{6}$ & $0.76$ & $jj$ \\
$e^+e^- \to cc H(\to ZZ)$ No Inv. & $1.20 \times 10^{6}$ & $0.59$ & $jj$ \\
$e^+e^- \to ss H(\to ZZ)$ No Inv. & $1.20 \times 10^{6}$ & $0.76$ & $jj$ \\
$e^+e^- \to \nu\bar{\nu} H(\to bb)$ & $1.20 \times 10^{6}$ & $26.7$ & $jj$ \\
$e^+e^- \to \nu\bar{\nu} H(\to cc)$ & $1.20 \times 10^{6}$ & $1.34$ & $jj$ \\
$e^+e^- \to \nu\bar{\nu} H(\to ss)$ & $1.20 \times 10^{6}$ & $1.1 \times 10^{-2}$ & $jj$ \\
$e^+e^- \to qq H(\to WW)$ & $1.10 \times 10^{6}$ & $11.5$ & $jj$ \\
$e^+e^- \to bb H(\to WW)$ & $1.00 \times 10^{6}$ & $6.45$ & $jj$ \\
$e^+e^- \to cc H(\to WW)$ & $1.20 \times 10^{6}$ & $5.02$ & $jj$ \\
$e^+e^- \to ss H(\to WW)$ & $1.20 \times 10^{6}$ & $6.45$ & $jj$ \\
\bottomrule
\end{tabular}
\end{table*}

The simulation of all backgrounds except $ZZ$, $Z(qq)$ and $WW$ is carried out using \whizard\cite{Kilian:2007gr} followed by parton shower and hadronisation implemented in \pythia\cite{Sjostrand:2006za}. The $ZZ$, $Z(qq)$, and $WW$ backgrounds are generated via \pythiaN\cite{Sjostrand:2014zea}. All the simulated events are processed through the parametric response of the proposed \textsc{IDEA} detector~\cite{IDEAStudyGroup:2025gbt} achieved through its implementation within the \textsc{Delphes} software~\cite{deFavereau:2013fsa}. The simulations have also implemented initial state radiation, final state radiation and a gaussian beam spread. 

A right-handed coordinate system is defined with its origin at the collision point. The $x$-axis points toward the enter of the FCC-ee collider, the $y$-axis is oriented vertically upward,  and the $z$-axis is aligned along the beam direction.  Azimuthal angles are measured from the $x$-axis, while polar angles are defined relative to the $z$-axis. For the IDEA detector simulation, electrons and muons with transverse momentum $p_T > 100~\rm{MeV}$  and pseudorapidity $|\eta| < 2.56$ are assumed to be fully reconstructed with 100\% efficiency.  After accounting for detector effects, including material-induced scattering, the identification efficiency for electrons and muons with energy $E > 2~\rm{GeV}$  and $|\eta| < 3$ is assumed to be 99\%.

The exclusive Durham $k_T$ algorithm implemented in the \textsc{FastJet} software~\cite{Ellis:1993tq,Catani:1991hj,Catani:1993hr,Cacciari:2011ma,fccjet} is used for jet clustering. Here, the distance between two particles is given by: 
\begin{equation}
    d_{ij} = 2~ \mathrm{min} (E_{i}^2,E_{j}^2) (1-\cos\theta_{ij})
\end{equation}

Prior to jet clustering, the hard electrons or muons are removed from the list of physics objects passed on to the jet clustering stage. The $N_{\text{jet}}$ parameter of the algorithm is set to two, hence reclustering objects entering the algorithm into two jets in the final state. We have used the $E$-recombination scheme. This algorithm also gives one access to  variables such the $d_{23}, d_{34}$ variables which are useful in discriminating events consistent with the two jets hypothesis versus events that may contain more jets but are reclustered as two jets. 

The Event generation and simulation software chain have been implemented within the \textsc{Key4HEP} software~\cite{Key4hep:2023nmr} stack.

\section{Analysis}\label{sec:analysis}

\begin{figure*}[htbp]
    \centering
    \includegraphics[width=0.45\linewidth]{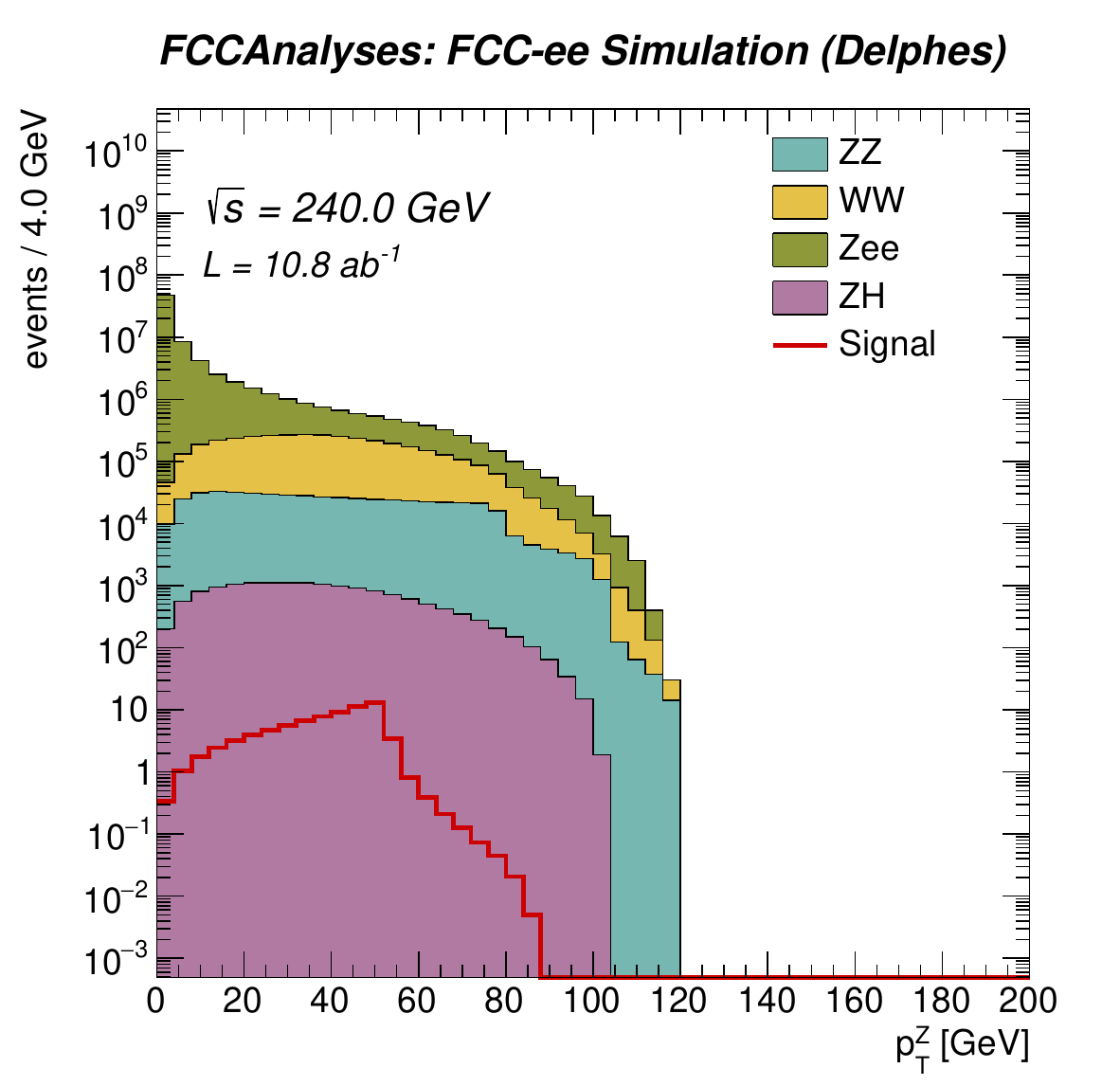}
    \includegraphics[width=0.45\linewidth]{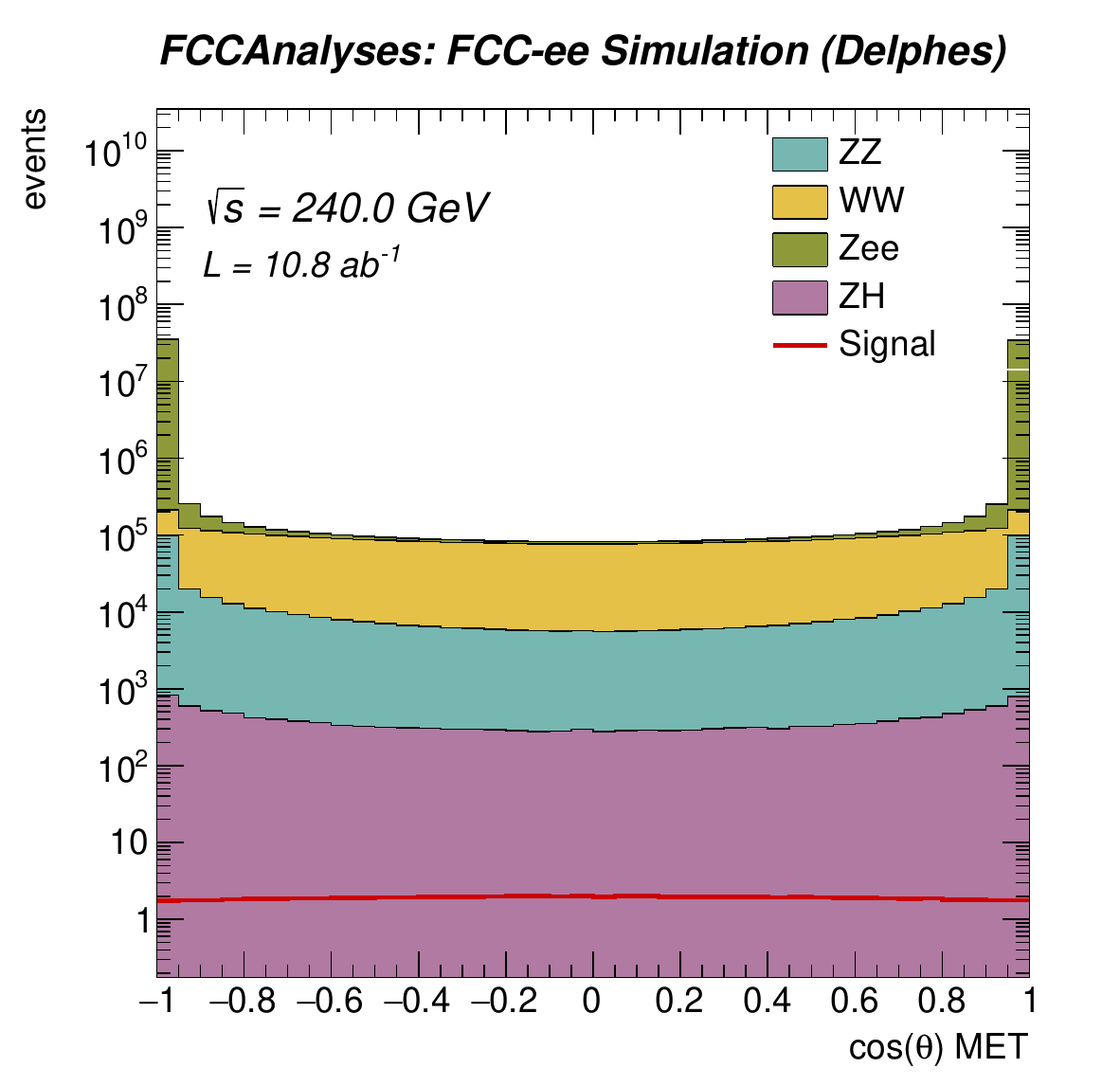}\\
    \includegraphics[width=0.45\linewidth]{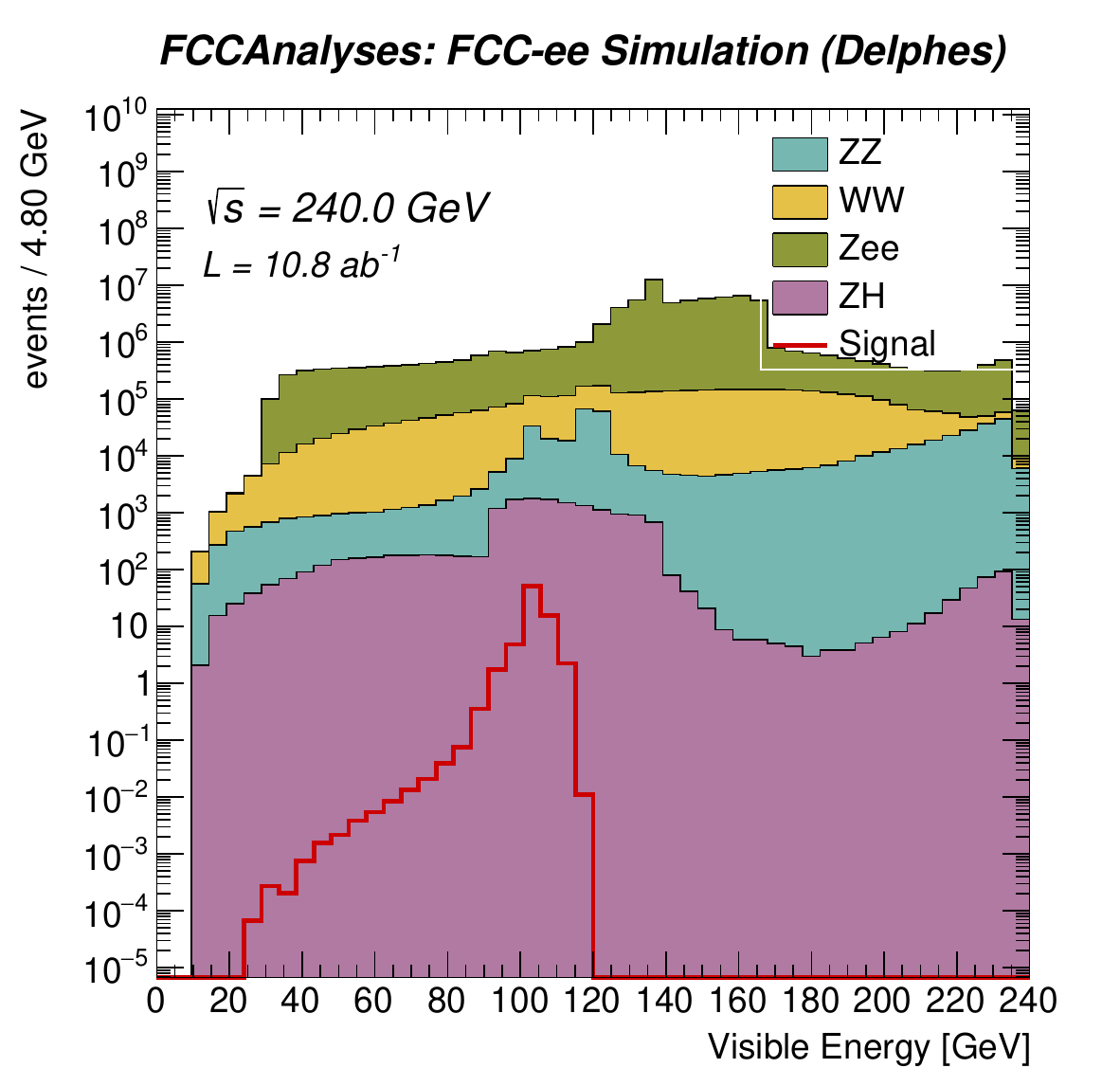}
    \includegraphics[width=0.45\linewidth]{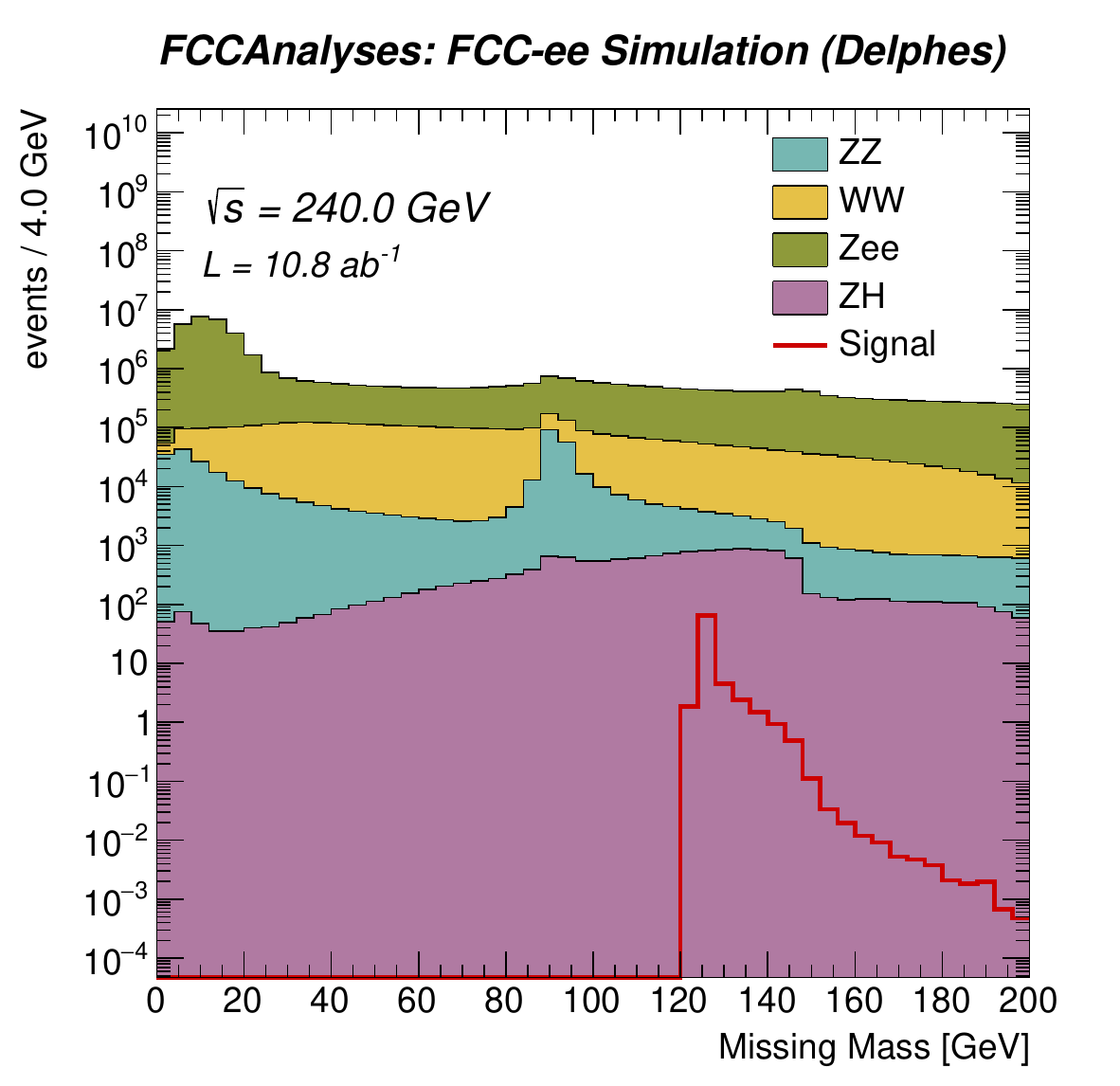}
    \caption{Kinematic distributions in the $Z(ee)H(inv.)$ channel after preselection criteria are applied.}
    \label{fig:ee_preselect}
\end{figure*}

\begin{figure*}[htbp]
    \centering
    \includegraphics[width=0.45\linewidth]{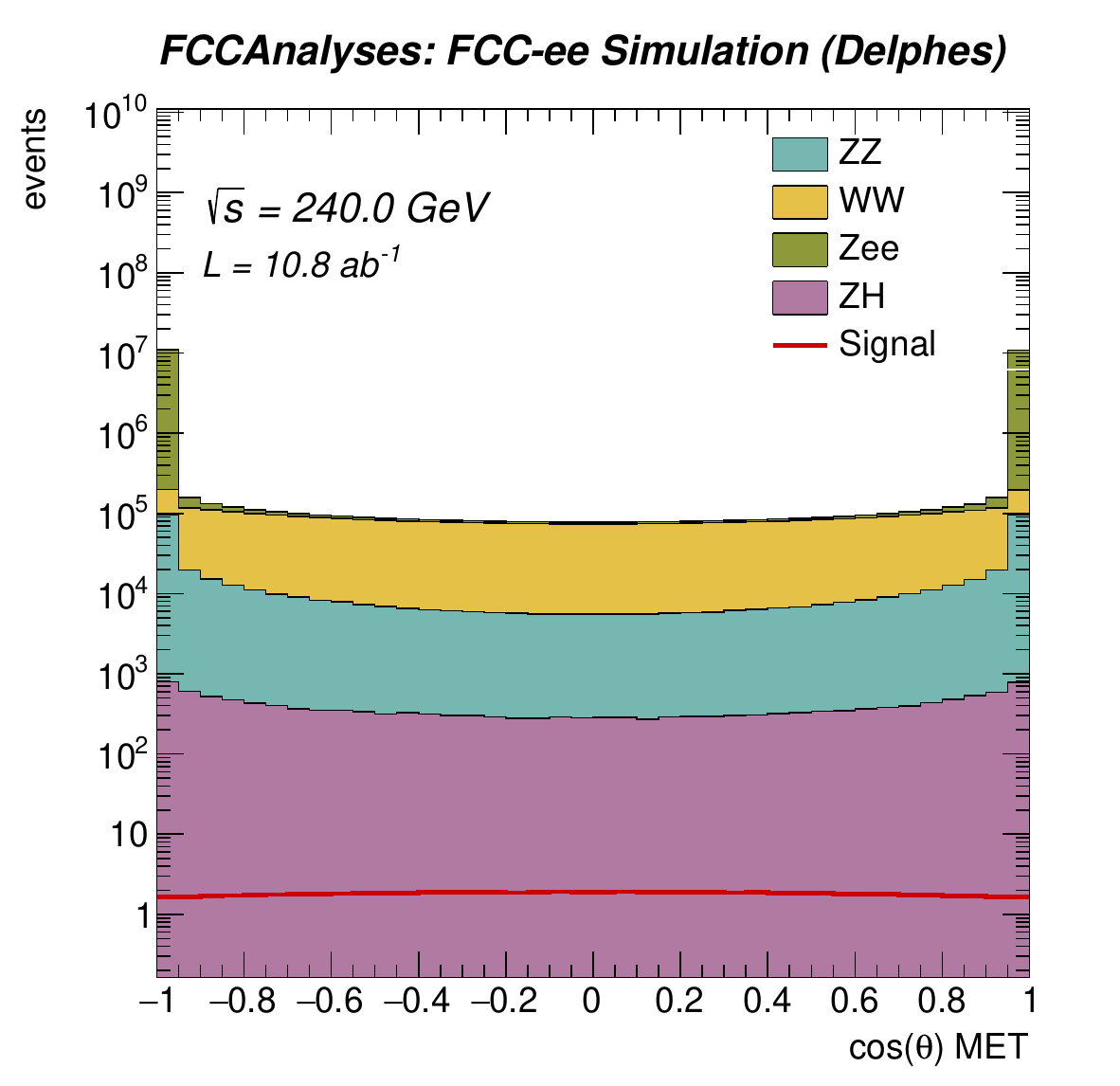}
    \includegraphics[width=0.45\linewidth]{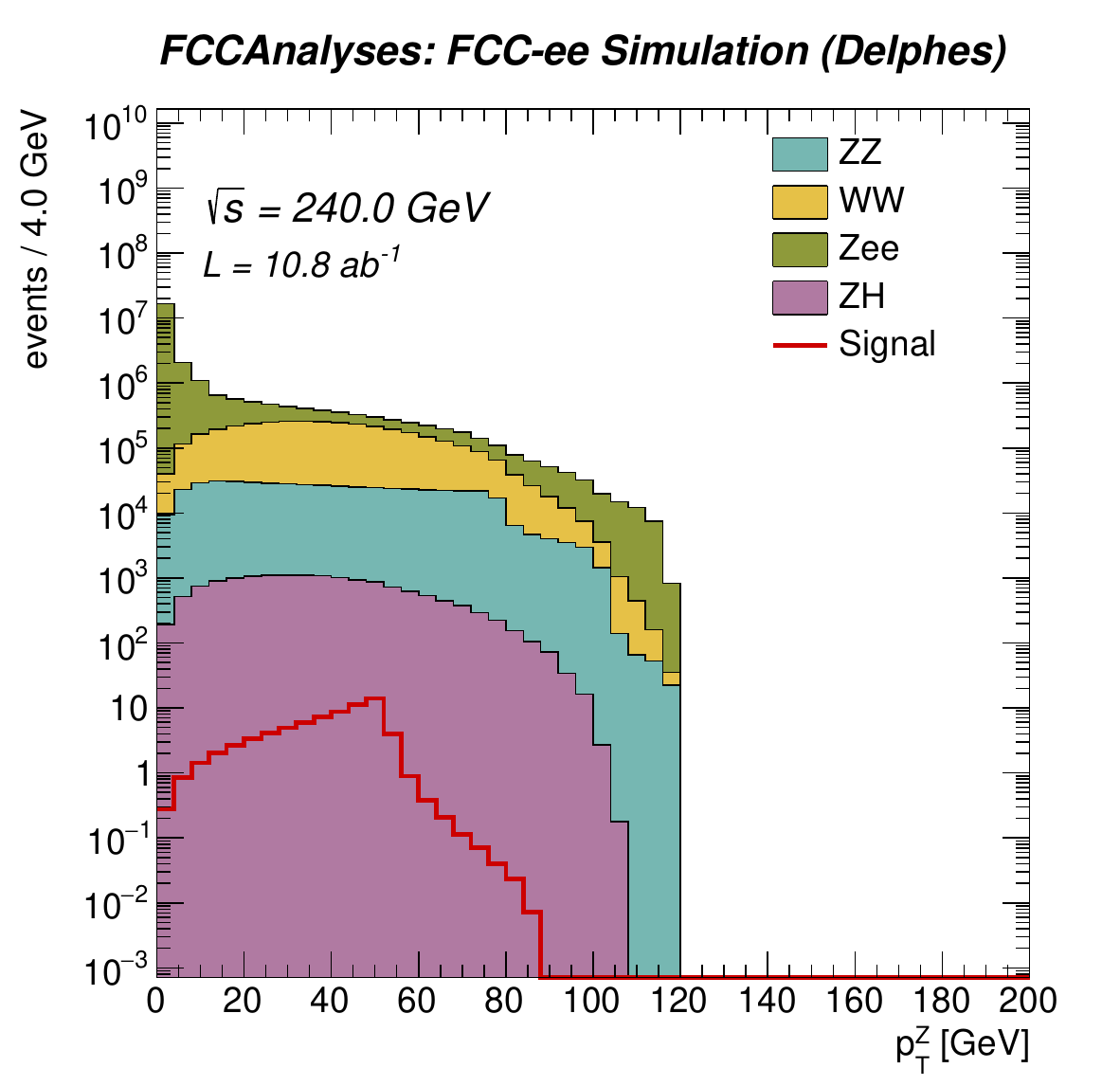}\\
    \includegraphics[width=0.45\linewidth]{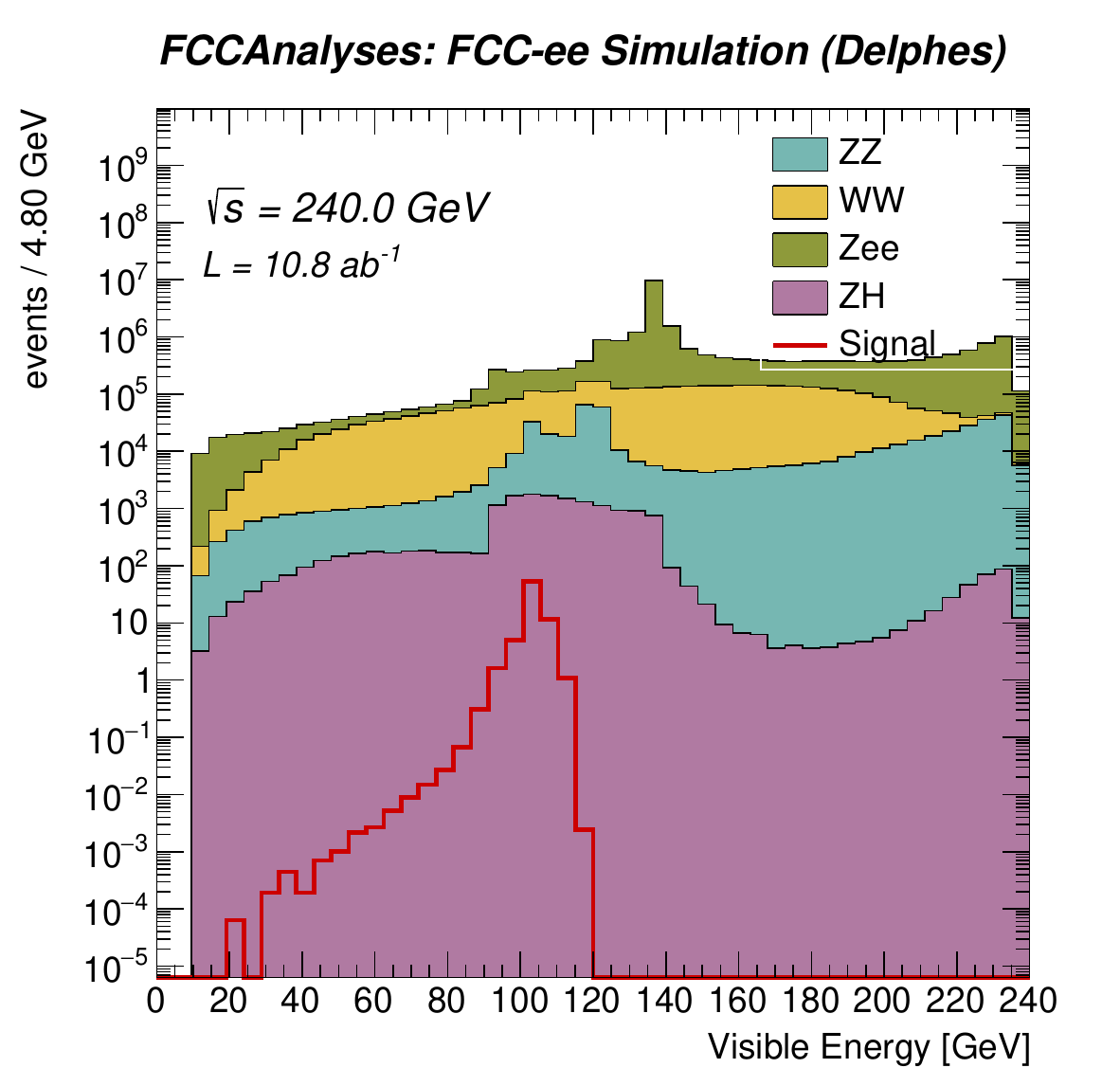}
    \includegraphics[width=0.45\linewidth]{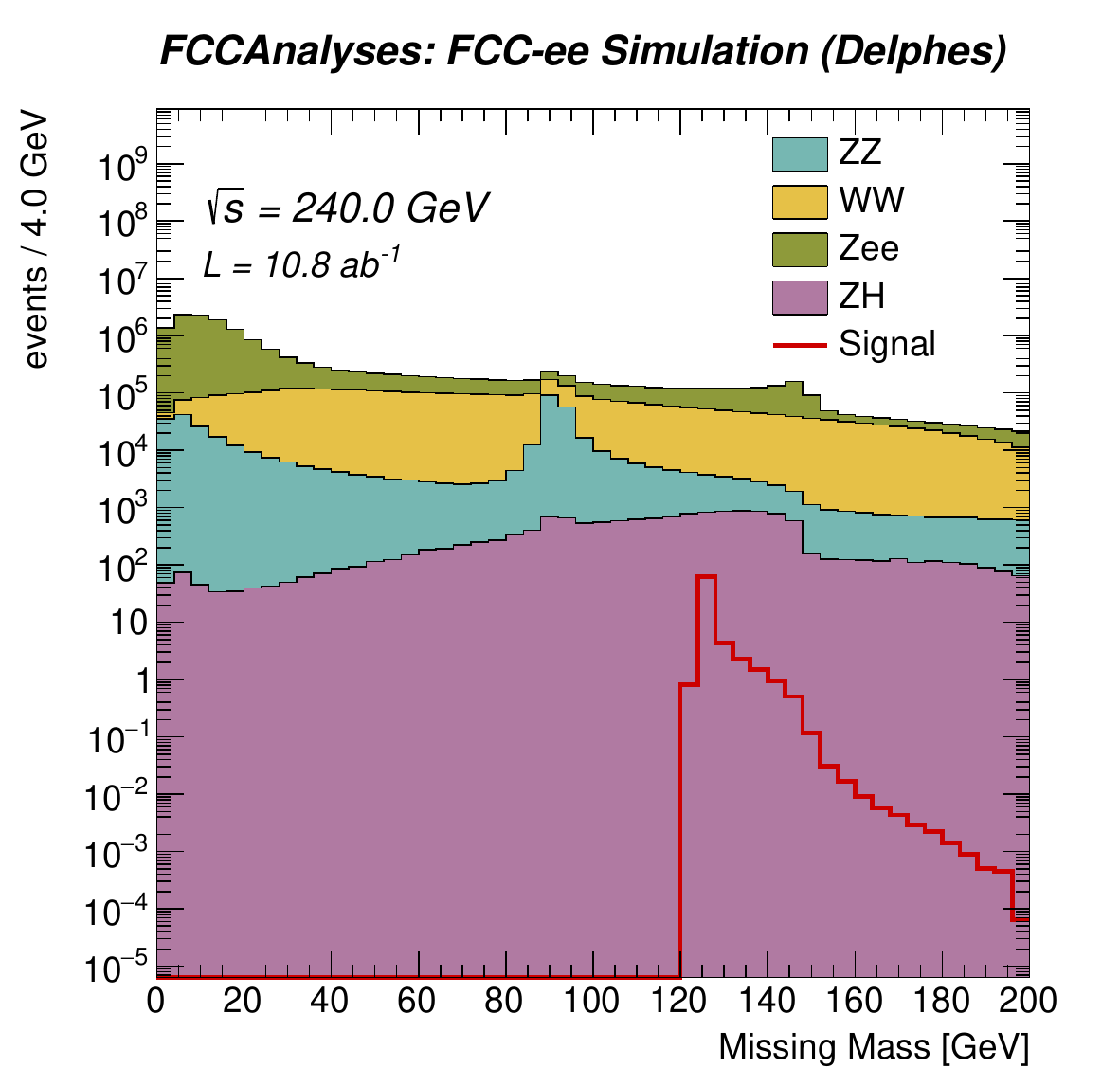}
    \caption{Kinematic distributions in the $Z(\mu\mu)H(inv.)$ channel after preselection criteria are applied.}
    \label{fig:mumu_preselect}
\end{figure*}

\begin{figure*}[htbp]
    \centering
    \includegraphics[width=0.45\linewidth]{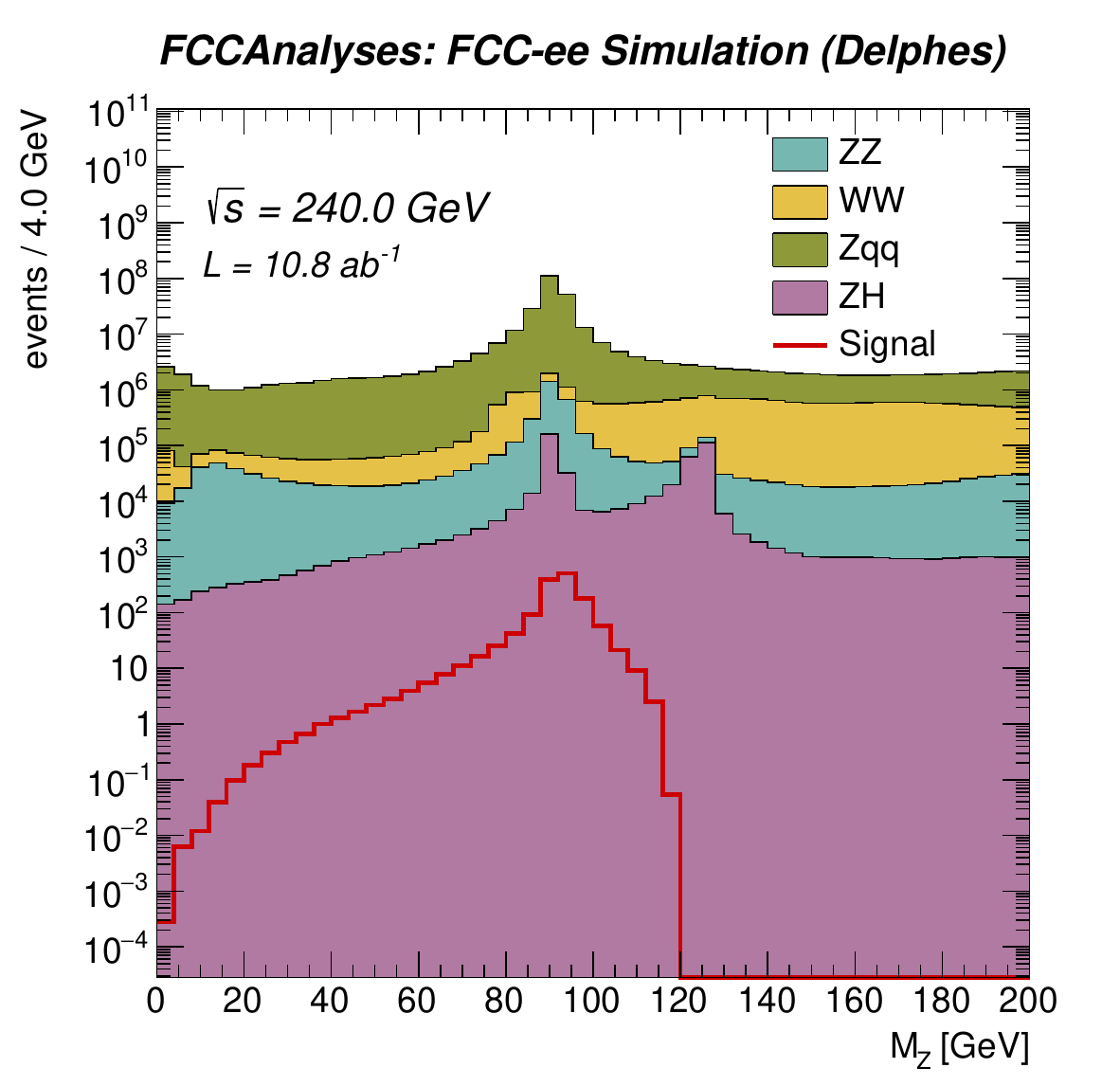}
    \includegraphics[width=0.45\linewidth]{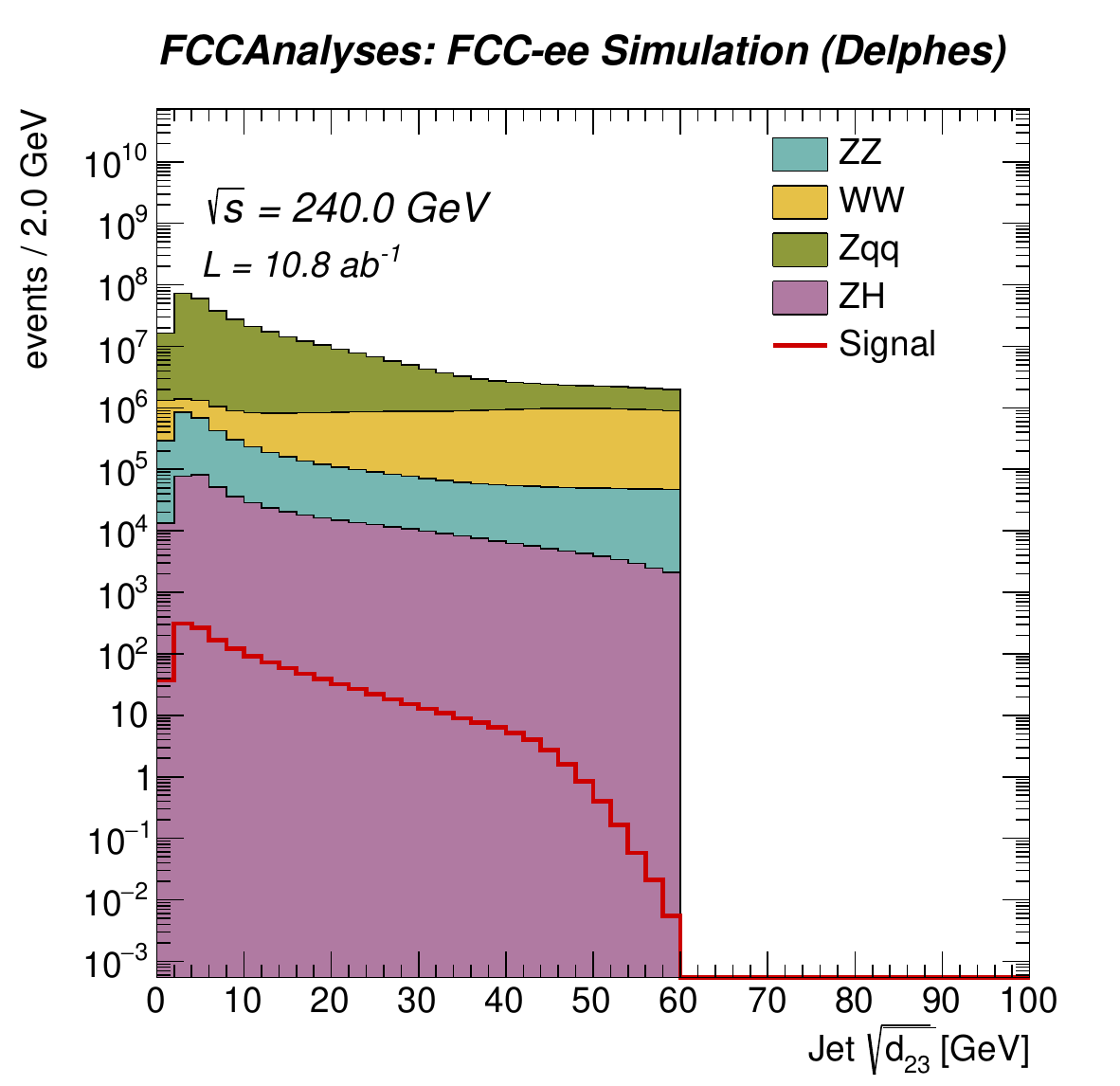}\\
    \includegraphics[width=0.45\linewidth]{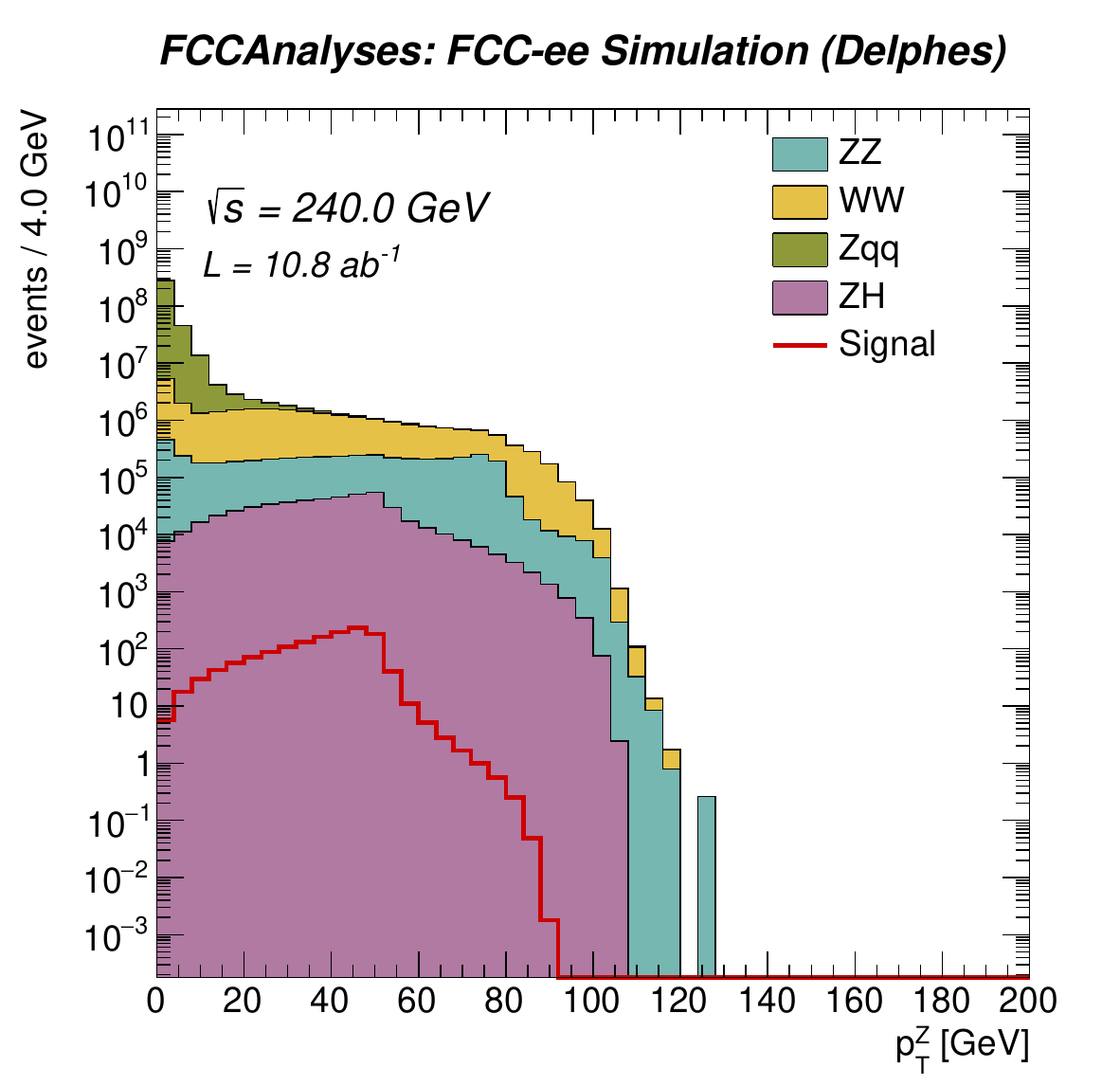}
    \includegraphics[width=0.45\linewidth]{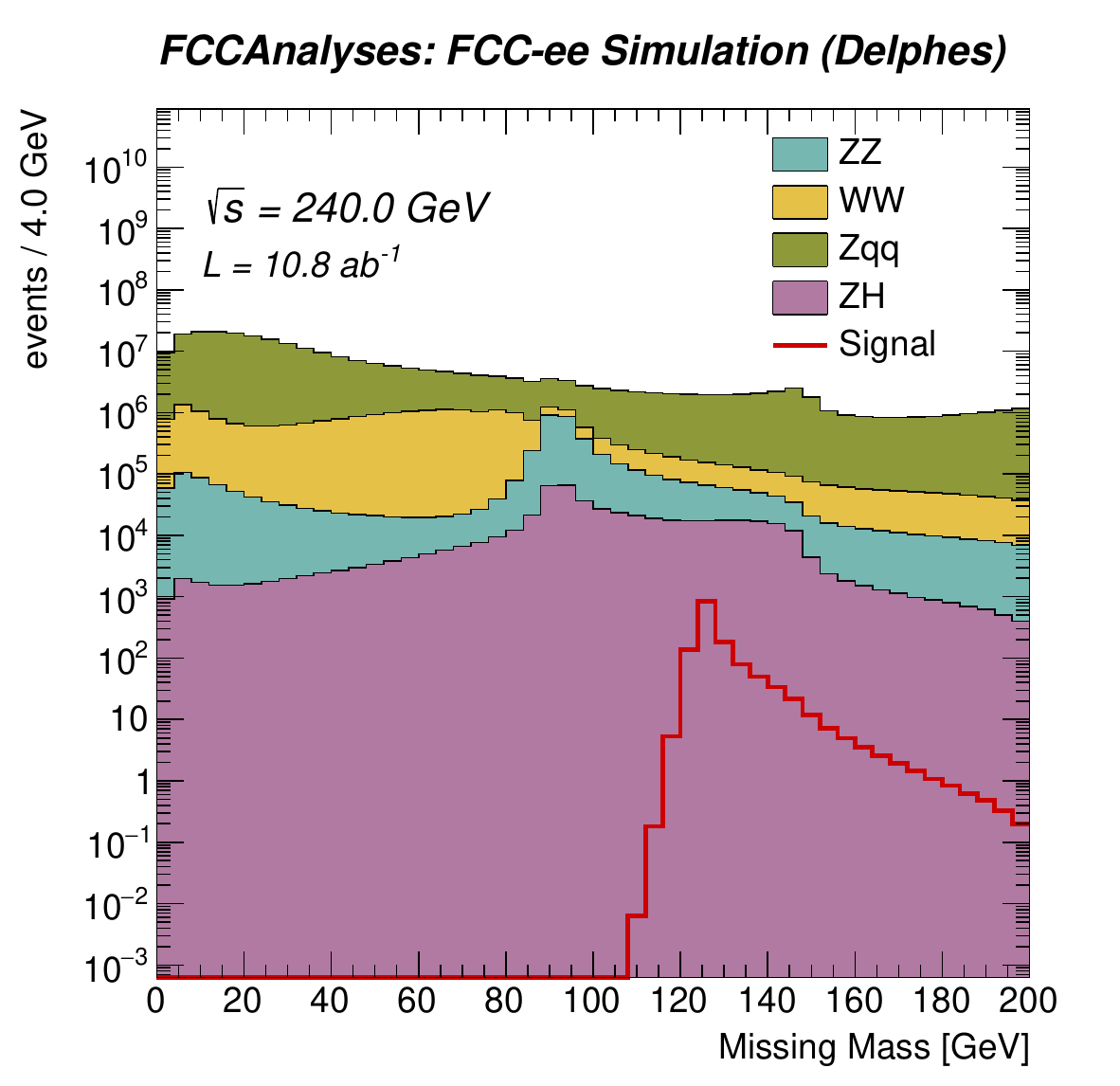}
    \caption{Kinematic distributions in the $Z(jj)H(inv.)$ channel after preselection criteria are applied.}
    \label{fig:qq_preselect}
\end{figure*}

The analysis is carried out in the \textsc{FCCAnalyses} software framework~\cite{helsens_2025_15528870}. In this section we discuss the analysis which we implemented in three orthogonal channels according to the decay mode of the $Z$ boson. In particular, we consider the decay modes with $Z$ boson decaying leptonically to electrons or muons and hadronic decays. The orthogonalization of the subanalyses is achieved at the preselection stage. In particular, for analysis with final states consisting of electrons (muons), we require exactly two electrons (muons) and zero muons (electrons) with their momenta $p > 5$ GeV. Additionally, the sum of charges is required to be zero. For the hadronic final states, the events are chosen such that the final states consisting of two jets constructed using the exclusive Durham $k_T$ algorithm as described in the previous section are chosen with zero electrons or muons with $p> 5$ GeV. Owing to the nature of invisible Higgs decay, we preselect events with missing momenta $p^{\rm miss} > 5$ GeV. Moreover, in the case of final state consisting of jets, we require $\sqrt{d_{23}} < 60$ GeV as well as $\sqrt{d_{34}} < 40$ GeV. The kinematic distributions for the final states after preselection stage are given in \autoref{fig:ee_preselect} for $Z(ee)H(inv.)$, \autoref{fig:mumu_preselect} for $Z(\mu\mu)H(inv.)$, and \autoref{fig:qq_preselect} for $Z(jj)H(inv.)$.

In each of the following sub-analyses, the $Z$ boson is reconstructed using its visible decay products. As we require the presence of exactly two electrons (muons) in the case of leptonically decaying $Z$ boson and since we have two jets in hadronic final state owing to the configuration of the jet clustering algorithm, the reconstruction procedure is simplified. The recoil mass of Higgs boson is computed using the visible decay products of the $Z$ boson. The recoil mass is defined as:

\begin{equation}
M_{\mathrm Recoil} = s + m_Z^2 - 2 E_z \sqrt{s}
\end{equation}

This reconstruction of variables is followed by a selection-based analysis. The selection criteria applied to the events is independent of the final states and is as follows. Firstly, we restrict the $Z$ boson mass to lie within $[87, 95]$ GeV. This ensures that events such as the ones coming from $WW$ background are suppressed. In order to veto events not consisting the Higgs boson, we impose requirement on the recoil mass such that it satisfies $120 ~\rm{GeV}<  m_{\text{recoil}} < 130$ GeV. This selection removes most of the events coming from $ZZ$, and from $Z\to f\bar{f}$ ($f$ can be a lepton or jet). The presence of neutrinos in the final state leads to signatures of missing four-momenta and angles of the missing momenta vector. The restriction is imposed on $|\cos\theta_{\text{miss}}| < 0.95$. In particular, $|\cos\theta_{\text{miss}}|$ variable allows us to veto a significant fraction of events coming from $Z\to f\bar{f}$.

For imposing each selection criteria we compute the significance achieved. This quantity is defined as $Z = S\sqrt{S+B}$ where $S, B$ represent the Signal and Background yield. The cutflow tables are presented in \autoref{tab:cut_ee} for $Z(ee)H(inv.)$, \autoref{tab:cut_mumu} for $Z(\mu\mu)H(inv.)$, and \autoref{tab:cut_qq} for $Z(jj)H(inv.)$.

The final selection criteria applied in the cutflow tables is the Multivariate selection. To construct this variable, we processed the preselected events through a Boosted Decision Tree (BDT) implemented in \textsc{XGBoost}~\cite{Chen:2016:XGBoost}. This allows one to construct a single discrimination that can help in separating the signal from background. 

The inputs to the Boosted Decision Tree (BDT) consists of a set of kinematic observables. These include the four-momenta of the $Z$-boson decay products, the missing four-momentum, transverse missing momentum, missing mass, and the total visible energy in the event. 
In addition the variables related to the reconstructed Higgs boson such as recoil mass are used. Variables obtained upon reconstructing the $Z$ boson are also included, such as its energy, momentum, and transverse momentum. Moreover, to improve the discriminating power of the BDT, we augment the feature set with variables derived from combinations of the four vector inputs. The momentum imbalance ($\mathcal{I}_{e_1 e_2}$) is defined in the $Z(ee)H(inv.)$ channel as: 

\begin{equation}
\mathcal{I}_{e_1 e_2} = \frac{p_T^{e_1} - p_T^{e_2}}{p_T^{e_1} + p_T^{e_2}}
\end{equation}

Additionally, we also define the ratios of electron $p_T$ with the missing $p_T$ ($\not{p}_T$) as follows:

\begin{align}
\mathcal{R}_{p_{T}^{e_1} / \not{p}_T} &= \frac{p_T^{e_1}}{\not{p}_T}, \\
\mathcal{R}_{p_{T}^{e_2} / \not{p}_T} &= \frac{p_T^{e_2}}{\not{p}_T}.
\end{align}

and define the ratio of total electron energy over total electron momenta: 

\begin{equation}
\mathcal{R}_{E^{e} / p^{e}}= \frac{E^{\text{electron 1}} + E^{\text{electron 2}}}{\left|\vec{p}^{\,\text{electron 1}}| + |\vec{p}^{\,\text{electron 2}}\right|}
\end{equation}

Additionally, angular variables between visible objects are also used. Similar variables are also defined in the other channels.

\begin{figure*}[htbp]
    \centering
    
    \begin{subfigure}{0.45\textwidth}
        \centering
        \includegraphics[width=\linewidth]{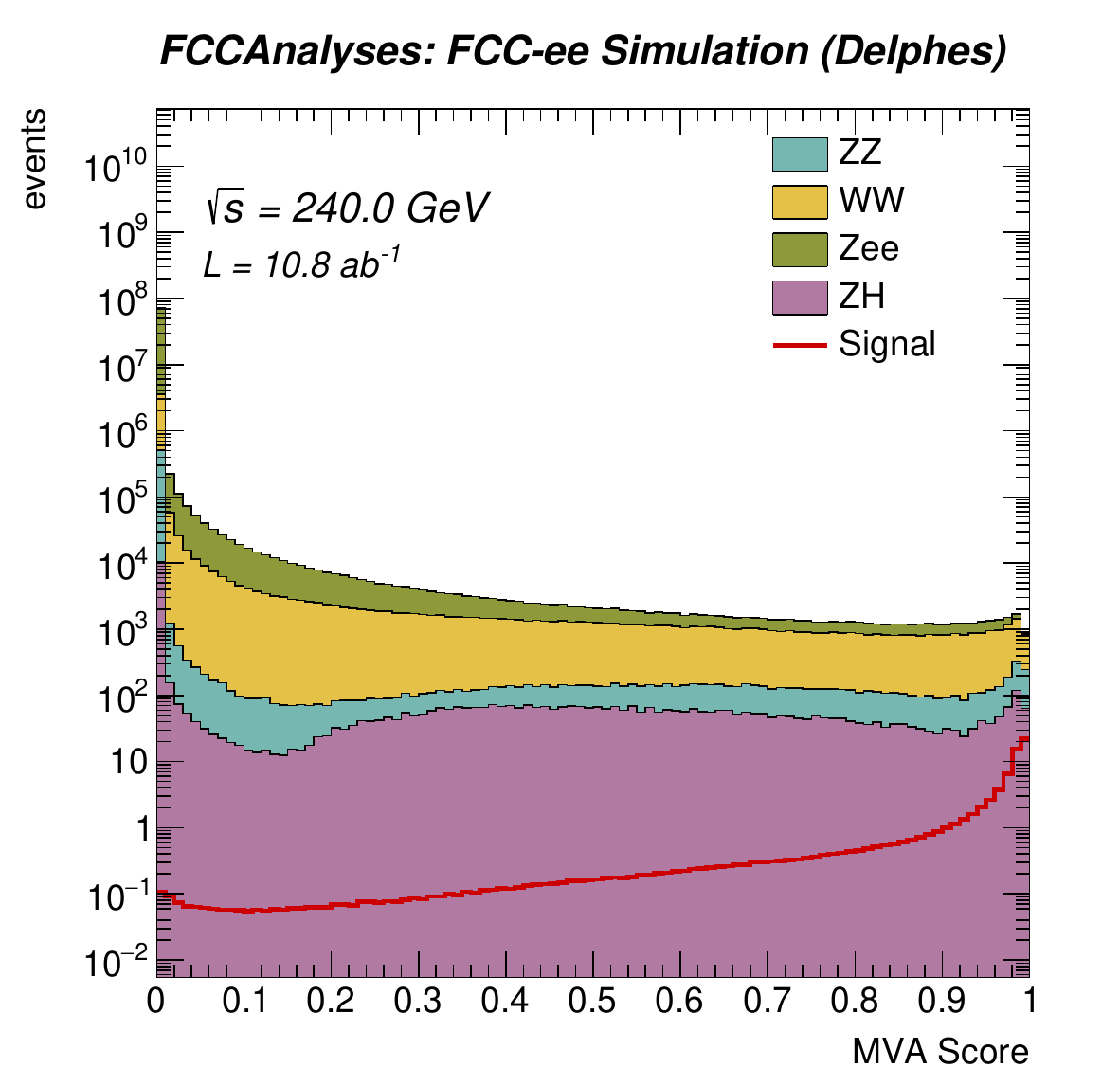}
        \caption{$ee$ channel}
        \label{fig:mva_ee}
    \end{subfigure}
    \hfill
    \begin{subfigure}{0.45\textwidth}
        \centering
        \includegraphics[width=\linewidth]{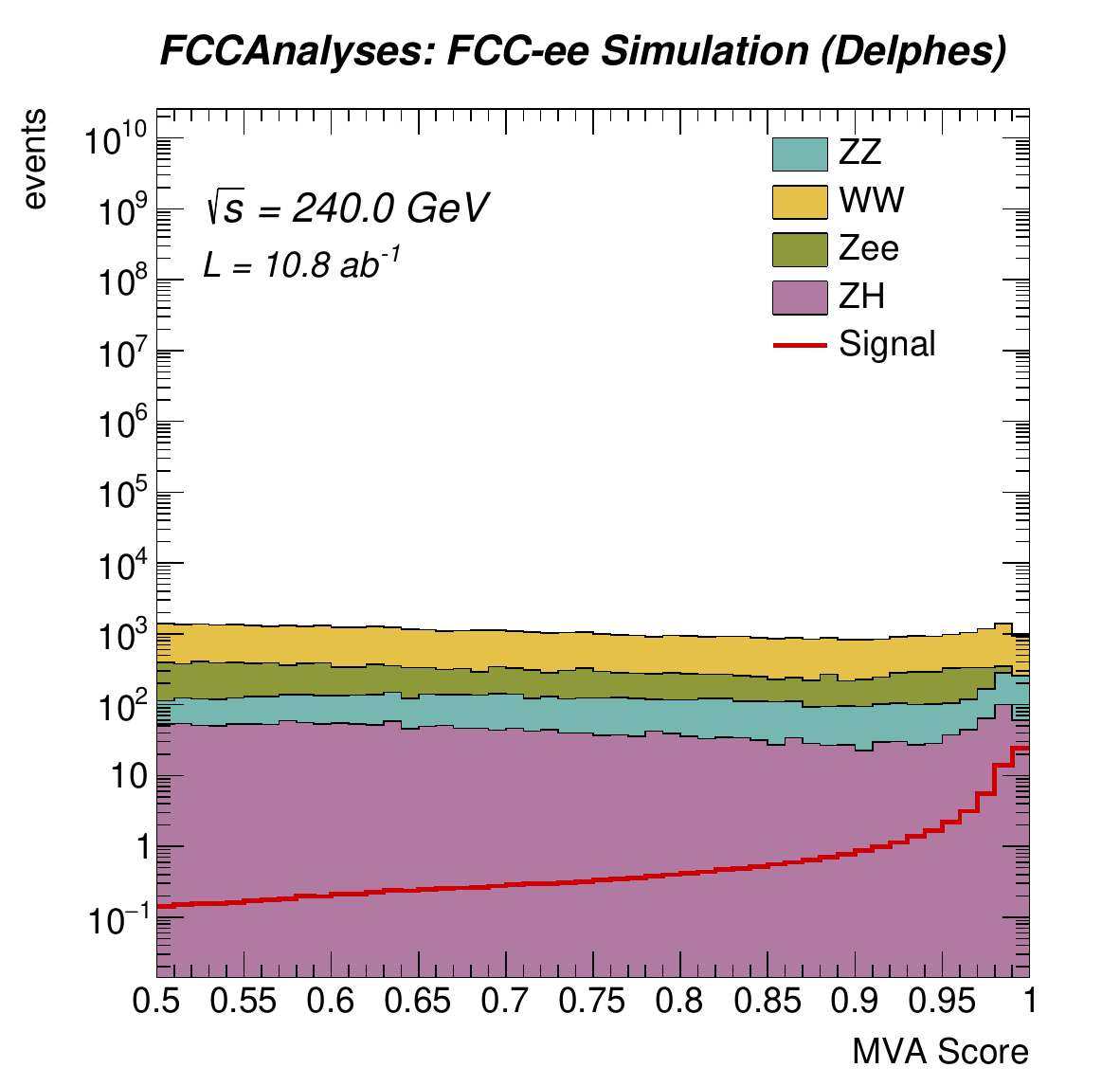}
        \caption{$\mu\mu$ channel}
        \label{fig:mva_mumu}
    \end{subfigure}
    \\
    \begin{subfigure}{0.45\textwidth}
        \centering
        \includegraphics[width=\linewidth]{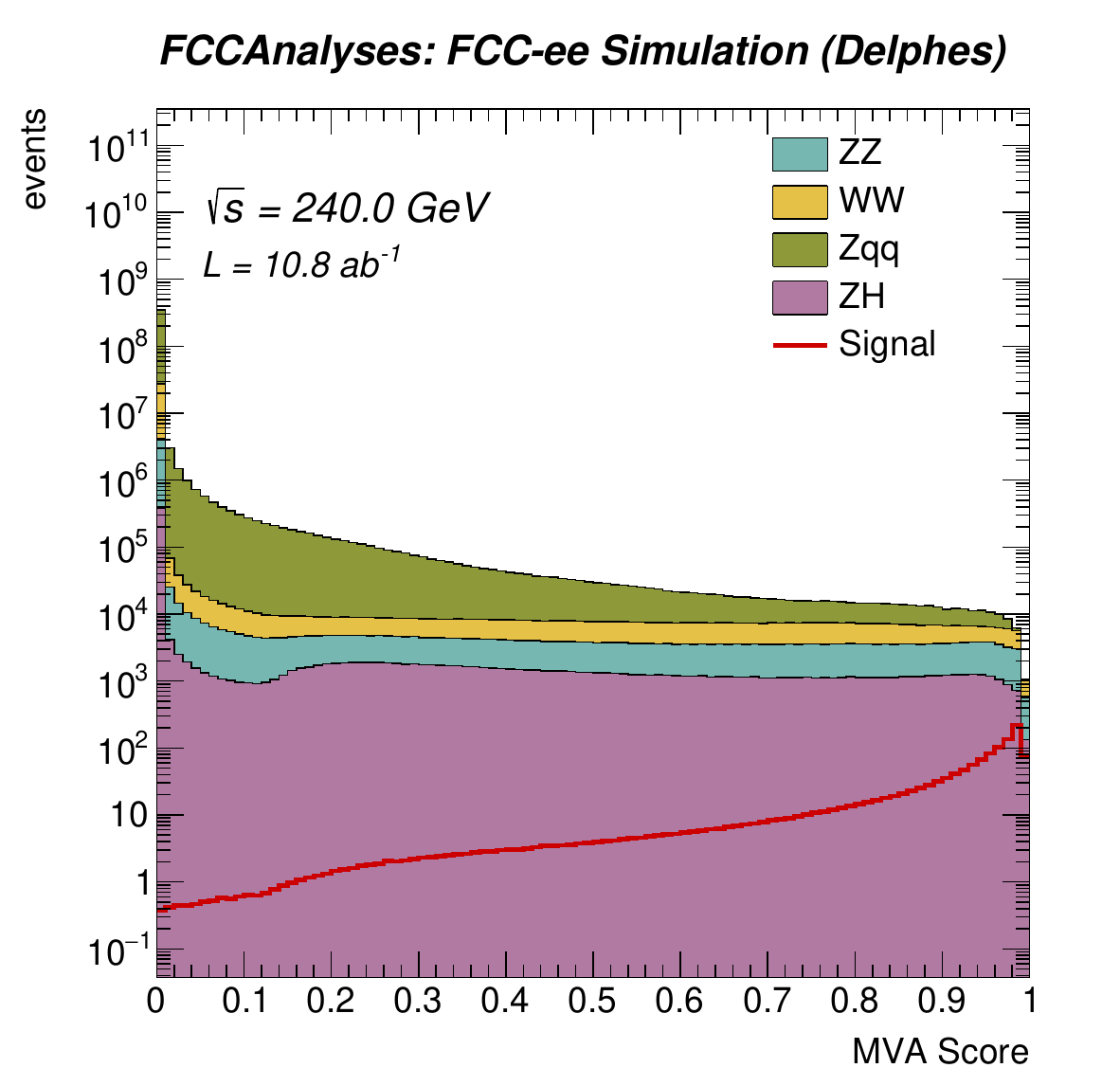}
        \caption{$qq$ channel}
        \label{fig:mva_qq}
    \end{subfigure}

    \caption{MVA score after preselection criteria for the $ee$, $\mu\mu$, and $qq$ channels.}
    \label{fig:mva}
\end{figure*}

The BDT used in this analysis is trained by using the hyperparameter setting as defined in \autoref{tab:hyper}.  The trained BDT model is then converted into ROOT~\cite{Brun:1997pa} format via \textsc{CERN Root Tmva}~\cite{Voss:2007jxm} for a streamlined inference within the \textsc{FCCAnalyses} software. The outcome of this study is a single MVA score that is used to discriminate the signal against the background. The distribution of the MVA score is given in \autoref{fig:mva_ee} for $Z(ee)H(inv.)$, \autoref{fig:mva_mumu} for $Z(\mu\mu)H(inv.)$, and \autoref{fig:mva_qq} for $Z(jj)H(inv.)$.  

Final selection is then made on the MVA score. We select events with their \text{MVA} score greater than 0.95. Thus we see a final significance of $0.6\sigma$ in both of the leptonic $Z$ decay channels, i.e., the $ee$ and $\mu\mu$ channels, whereas in the hadronic $Z$ decay channel a significance of $3.1\sigma$ is achieved.

\begin{figure*}[htbp]
    \centering
    
    \begin{subfigure}{0.7\textwidth}
        \centering
        \includegraphics[width=\linewidth]{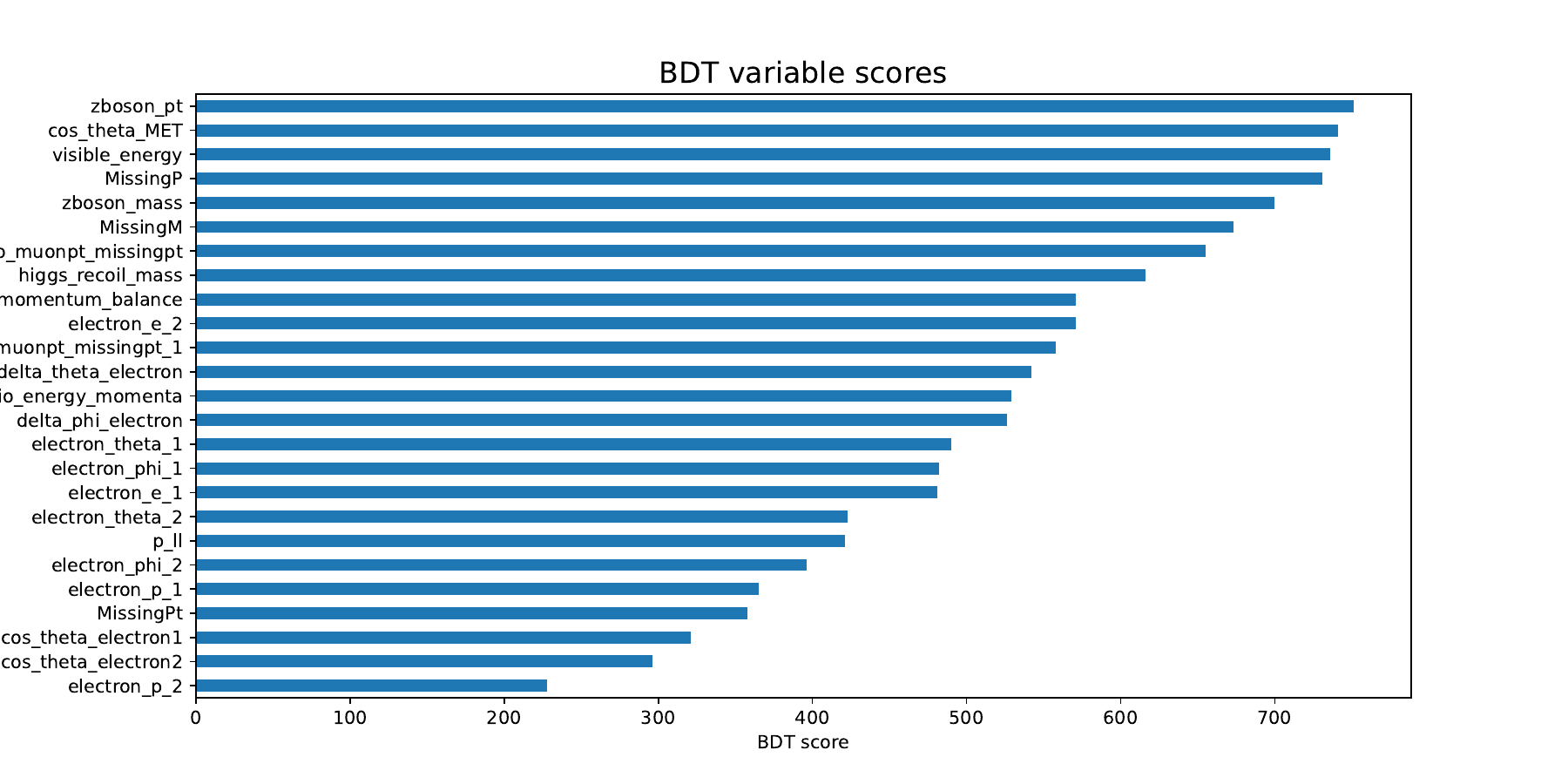}
        \caption{$ee$ channel}
        \label{fig:imp_ee}
    \end{subfigure}
    \hfill
    \begin{subfigure}{0.7\textwidth}
        \centering
        \includegraphics[width=\linewidth]{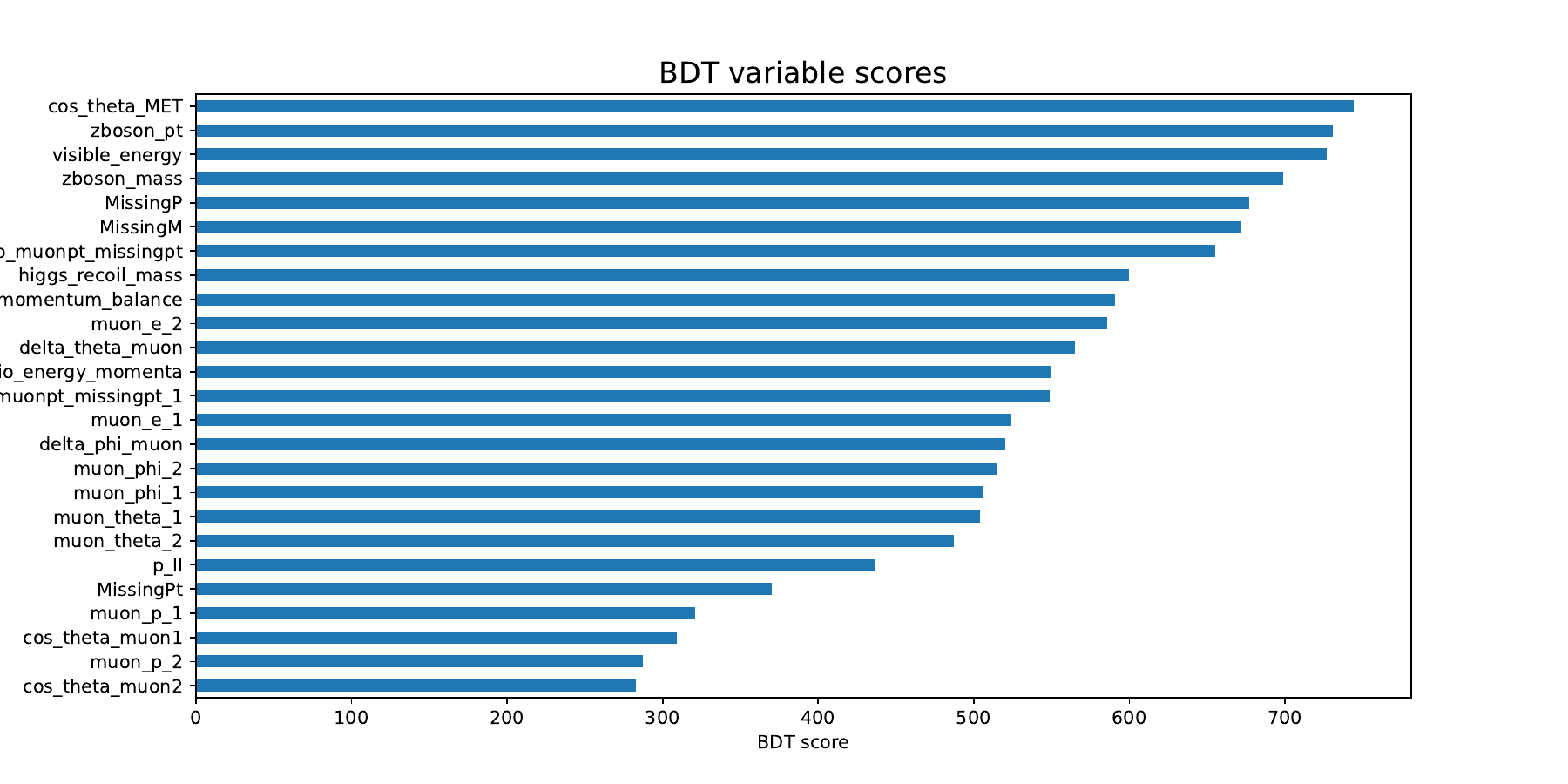}
        \caption{$\mu\mu$ channel}
        \label{fig:imp_mumu}
    \end{subfigure}
    \\
    \begin{subfigure}{0.7\textwidth}
        \centering
        \includegraphics[width=\linewidth]{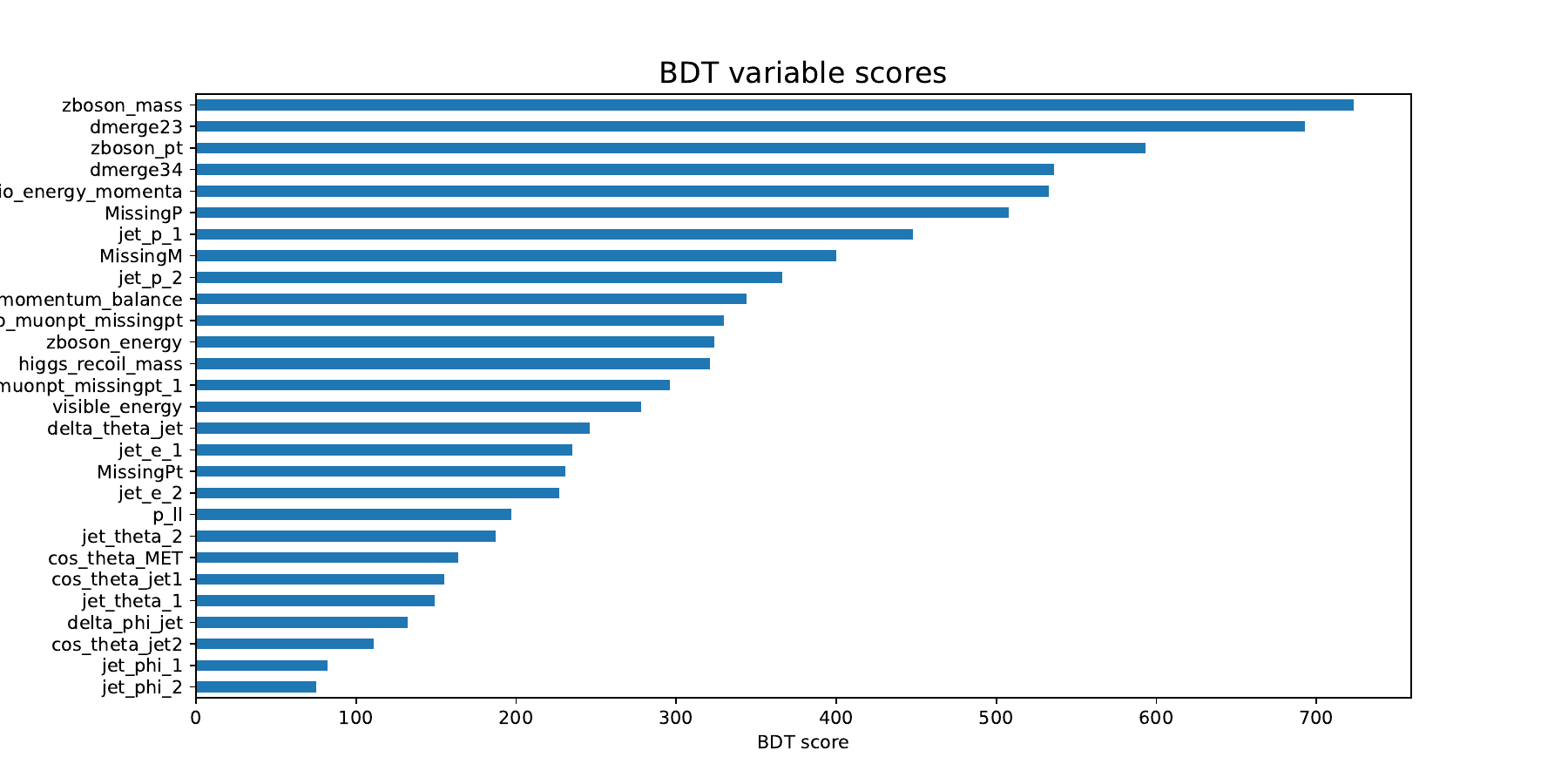}
        \caption{$qq$ channel}
        \label{fig:imp_qq}
    \end{subfigure}

\caption{Ranking of input variables by their importance in the BDT for the $ee$, $\mu\mu$, and $qq$ channels.}
\label{fig:var_imp}
\end{figure*}

Based on MVA distribution shown in \autoref{fig:mva}, a reduction in the number of background events is observed in region  $MVA > 0.9$ for the $Z(jj)H(inv.)$ while no such feature is seen in the $Z(\ell\ell)H(inv.)$ channels. This behavior may be understood in terms of the relative importance of input variables used in training the BDT models for the three channels, as given in \autoref{fig:var_imp}. Accordingly, it is found that the leptonic channels rank the $p_T^{Z}$, $\cos(\theta_{\rm miss})$ and $E^{\rm{ vis}}$ as the top three important variables whereas in the hadronic channel, the $M_{Z}$, $d_{23}$ and $p_T^{Z}$ ranked as the highest three variables based on their importance. The corresponding kinematic distributions are presented in \autoref{fig:ee_preselect}, \autoref{fig:mumu_preselect}, and \autoref{fig:qq_preselect}. We note that $p_T^{Z}$ is among the most important variables across all channels. However, the $\cos(\theta_{\rm miss})$ distribution which is ranked highest by the leptonic channel has a relatively flat shape which leads weaker discriminating power compared with the $d_{23}$ distribution in $Z(jj)H(inv.)$ channel. This effect combined with the larger branching fraction of the $Z$ boson decay to hadrons ($\mathcal{B}(Z\to \rm{hadronic}) \sim 69\%$) compared with $\mathcal{B}(Z\to \mu\mu/ee) \sim 3.4\%$) leads to higher sensitivity in the $Z(jj)H(inv.)$ channel compared to the $Z(\ell\ell)H(inv.)$ channels.

\begin{table*}[t]
\centering
\small
\caption{Cut-flow table for the $Z(ee)H(\text{inv})$ channel showing signal ($S$), background contributions, total background, and statistical significance.}
\begin{tabular}{lccccccc}
\toprule
Selection & $S$ & ZZ & WW & $Zee$ & ZH (bkg) & $B_{\text{tot}}$ & $Z=\frac{S}{\sqrt{S+B}}$ \\
\midrule
Preselection & 76 & $5.08\times10^{5}$ & $3.32\times10^{6}$ & $7.04\times10^{7}$ & $1.53\times10^{4}$ & $7.42\times10^{7}$ & 0.01 \\
$87<m_Z<95$ GeV & 51 & $1.95\times10^{5}$ & $1.29\times10^{5}$ & $1.11\times10^{7}$ & $9.73\times10^{3}$ & $1.15\times10^{7}$ & 0.02 \\
$120<m_{\text{recoil}}<130$ GeV & 42 & $1.22\times10^{4}$ & $1.95\times10^{4}$ & $3.00\times10^{5}$ & $1.71\times10^{3}$ & $3.34\times10^{5}$ & 0.07 \\
$ |\cos\theta_{\text{miss}}| < 0.80$ & 34 & $3.34\times10^{3}$ & $1.46\times10^{4}$ & 12 & $1.22\times10^{3}$ & $1.91\times10^{4}$ & 0.24 \\
$\text{Ratio } p_{T}^{e}/p_{T}^{\text{miss}} < 2.0$ & 34 & $2.84\times10^{3}$ & $1.43\times10^{4}$ & 0 & $1.16\times10^{3}$ & $1.83\times10^{4}$ & 0.25 \\
$120 < m_{\text{miss}} < 140$ GeV & 34 & $1.69\times10^{3}$ & $9.04\times10^{3}$ & 0 & $9.78\times10^{2}$ & $1.17\times10^{4}$ & 0.31 \\
$\text{MVA} > 0.95$ & 29 & 388 & $1.82\times10^{3}$ & 0 & 216 & $2.42\times10^{3}$ & 0.59 \\
\bottomrule
\end{tabular}
\label{tab:cut_ee}
\end{table*}

\begin{table*}[t]
\centering
\small
\caption{Cut-flow table for the $Z(\mu\mu)H(\text{inv})$ channel showing signal ($S$), background contributions, total background, and statistical significance.}
\begin{tabular}{lccccccc}
\toprule
Selection & $S$ & ZZ & WW & $Z\mu\mu$ & ZH (bkg) & $B_{\text{tot}}$ & $Z=\frac{S}{\sqrt{S+B}}$ \\
\midrule
Preselection & 73 & $5.01\times10^{5}$ & $3.17\times10^{6}$ & $2.21\times10^{7}$ & $1.53\times10^{4}$ & $2.58\times10^{7}$ & 0.01 \\
$87<m_Z<95$ GeV & 55 & $2.13\times10^{5}$ & $1.25\times10^{5}$ & $1.03\times10^{7}$ & $1.08\times10^{4}$ & $1.07\times10^{7}$ & 0.02 \\
$120<m_{\text{recoil}}<130$ GeV & 46 & $1.30\times10^{4}$ & $1.83\times10^{4}$ & $1.63\times10^{5}$ & $1.92\times10^{3}$ & $1.97\times10^{5}$ & 0.10 \\
$ |\cos\theta_{\text{miss}}| < 0.80$ & 38 & $3.53\times10^{3}$ & $1.36\times10^{4}$ & 0 & $1.39\times10^{3}$ & $1.85\times10^{4}$ & 0.27 \\
$\text{Ratio } p_{T}^{\mu}/p_{T}^{\text{miss}} < 2.0$ & 38 & $3.00\times10^{3}$ & $1.34\times10^{4}$ & 0 & $1.31\times10^{3}$ & $1.77\times10^{4}$ & 0.28 \\
$120 < m_{\text{miss}} < 140$ GeV & 38 & $1.90\times10^{3}$ & $9.62\times10^{3}$ & 0 & $1.14\times10^{3}$ & $1.27\times10^{4}$ & 0.33 \\
$\text{MVA} > 0.95$ & 33 & 402 & $1.87\times10^{3}$ & 0 & 234 & $2.51\times10^{3}$ & 0.65 \\
\bottomrule
\end{tabular}
\label{tab:cut_mumu}
\end{table*}

\begin{table*}[t]
\centering
\small
\caption{Cut-flow table for the $Z(jj)H(\text{inv})$ channel showing signal ($S$), background contributions, total background, and statistical significance.}
\begin{tabular}{lccccccc}
\toprule
Selection & $S$ & ZZ & WW & $Zqq$ & ZH (bkg) & $B_{\text{tot}}$ & $Z=\frac{S}{\sqrt{S+B}}$ \\
\midrule
Preselection & $1.39\times10^{3}$ & $4.15\times10^{6}$ & $2.39\times10^{7}$ & $3.35\times10^{8}$ & $5.14\times10^{5}$ & $3.64\times10^{8}$ & 0.07 \\
$87<m_Z<95$ GeV & 850 & $1.93\times10^{6}$ & $1.04\times10^{6}$ & $1.64\times10^{8}$ & $1.98\times10^{5}$ & $1.68\times10^{8}$ & 0.06  \\
$120<m_{\text{recoil}}<130$ GeV & 702 & $5.46\times10^{4}$ & $3.77\times10^{4}$ & $1.75\times10^{6}$ & $3.26\times10^{4}$ & $1.88\times10^{6}$ & 0.51 \\
$ |\cos\theta_{\text{miss}}| < 0.80$ & 574 & $3.86\times10^{4}$ & $2.56\times10^{4}$ & 288 & $2.37\times10^{4}$ & $8.82\times10^{4}$ & 1.93 \\
$\text{MVA} > 0.95$ & 341 & $5.93\times10^{3}$ & $2.52\times10^{3}$ & 57 & $2.89\times10^{3}$ & $1.14\times10^{4}$ & 3.14 \\
\bottomrule
\end{tabular}
\label{tab:cut_qq}
\end{table*}

\begin{table*}[t]
\centering
\small
\caption{XGBoost hyperparameters used in the BDT training.}
\label{tab:hyper}
\begin{tabular}{ll ll}
\toprule
\textbf{Parameter} & \textbf{Value} & \textbf{Parameter} & \textbf{Value} \\
\midrule
Objective & binary:logistic & Eval metric & logloss \\
Max depth & 6 & Subsample & 0.5 \\
Colsample by tree & 0.5 & Tree method & hist \\
Number of estimators & 500 & Early stopping rounds & 30 \\
Learning rate & 0.15 & Gamma & 3 \\
Min child weight & 10 & Max delta step & 0 \\
\bottomrule
\end{tabular}
\end{table*}

\newpage
\clearpage

\section{Statistical Analysis}\label{sec:stat}

As the significance of $5\sigma$ is not achieved in any of the channels we derive the upper limits on the $\mathcal{B}(H\to inv.)$. The statistical analysis to evaluate the upper limits is performed using the \textsc{CMS Combine} software~\cite{CMS:2024onh}.  We use the missing mass $M_{\text{miss}}$ as an input variable for statistical analysis. The pseudo-data is obtained by combining the signal and background distributions after scaling these with their corresponding cross-section and luminosity.  A normalisation uncertainty of 10\% is assigned for the modeling of background processes. The luminosity is assigned an uncertainty of 1\%. The \textsc{CMS Combine} software's implementation of the \texttt{AsymptoticLimits} function which is based on Ref.~\cite{Cowan:2010js} is used to obtain the limits.  We find the following 95\% confidence level upper limits on  $\mathcal{B}(H\to inv.)$: 0.43\% in the $Z\to ee$ channel;  0.35\% in the $Z\to ee$ channel and 0.15\% in the $Z\to jj$ channel. The combined upper limit at 95\% C.L. is found as 0.15\%.  A summary of the results is given in \autoref{tab:upperlimits}.

\begin{table}[t]
\centering
\small
\caption{Expected 95\% CL upper limits on the branching fraction for the $ee$, $\mu\mu$, and $qq$ channels.}
\label{tab:upperlimits}
\renewcommand{\arraystretch}{1.3}  
\begin{tabular}{lc}
\toprule
\textbf{Physics} & \textbf{Upper Limit on $\mathcal{B}(H\to inv.)$} \\
\midrule
$Z(ee)H(\mathrm{inv})$      & $0.43\%$ \\
$Z(\mu\mu)H(\mathrm{inv})$  & $0.35\%$ \\
$Z(qq)H(\mathrm{inv})$      & $0.15\%$ \\
\textbf{Combined}                    & $0.15\%$ \\
\bottomrule
\end{tabular}
\end{table}

\section{Conclusions}

In this work, we studied the invisible decays of Higgs boson in the $ZH$ process at the FCCee at $\sqrt{s} = 240 \text{ GeV}$. We analyzed the  invisible Higgs boson decay in the three different physics processes: $Z(\to ee)H(\to inv)$, $Z(\to \mu\mu)H(\to inv)$,  and the $Z(\to jj)H(\to inv)$ channels. The $Z\to jj$ channel in this analysis consists of $Z\to bb$, $Z\to cc$, $Z\to ss$ and $Z\to qq$ channels which are combined with appropriate weighting with respect to their branching fraction.  We find upper limits on the $\mathcal{B}(H\to inv.)$.  Among the three channels, we note that the hadronic channel owing to its higher event yield ($\mathcal{B}(Z\to \rm{hadronic}) \sim 69\%$ versus $\mathcal{B}(Z\to \mu\mu/ee) \sim 3.4\%$), has higher sensitivity in deriving the upper limit for the  $H(\to inv)$ process.  A combined upper limit 95\% C.L. on the  $\mathcal{B}(H\to inv.)$ for three channels is found as $0.15\%$.

\bibliographystyle{JHEP}
\bibliography{biblio}

\end{document}